\documentclass[aip, amsmath,amssymb,reprint,nofootinbib]{revtex4-1}
\usepackage{graphicx}
\usepackage{hyperref}
\usepackage{ccaption}

\newcommand{\XU}{{X_\text{U}}}
\newcommand{\Dt}{dt}
\newcommand{\vf}{\mathbf{F}}

\begin{document}

\title{Stochastic unfolding of nanoconfined DNA: experiments, model and Bayesian analysis}

\author{Jens Krog} 
 \affiliation{MEMPHYS - Center for Biomembrane Physics, Department of Physics, Chemistry, and Pharmacy, University of Southern Denmark, Odense, Denmark}
\affiliation{Department of Astronomy and Theoretical Physics, Lund University,
  Lund, Sweden}

\author{Mohammadreza Alizadehheidari}
\affiliation{Department of Chemical and Biological Engineering, Chalmers University of Technology, Gothenburg, Sweden}

\author{Erik Werner} 
\affiliation{Department of Physics, Gothenburg University, Gothenburg, Sweden}

\author{Santosh Kumar Bikkarolla},
\affiliation{Department of Chemical and Biological Engineering, Chalmers
  University of Technology, Gothenburg, Sweden}

\author{Jonas O. Tegenfeldt}
\affiliation{Department of Physics and NanoLund, Lund University, Lund, Sweden}
 
\author{Bernhard Mehlig}
\affiliation{Department of Physics, Gothenburg University, Gothenburg, Sweden}

\author{Michael A. Lomholt} 
\affiliation{MEMPHYS - Center for Biomembrane Physics, Department of Physics, Chemistry, and Pharmacy, University of Southern Denmark, Odense, Denmark}
 
\author{Fredrik Westerlund}
\affiliation{Department of Chemical and Biological Engineering, Chalmers University of Technology, Gothenburg, Sweden}
 
\author{Tobias Ambj\"ornsson}
\email{tobias.ambjornsson@thep.lu.se}
\affiliation{Department of Astronomy and Theoretical Physics, Lund University, Lund, Sweden}

\date{\today}
\begin{abstract}
Nanochannels provide means for detailed experiments on the effect of
confinement on biomacromolecules, such as DNA.  We here introduce a model for
the complete unfolding of DNA from the circular to linear configuration. Two
main ingredients are the entropic unfolding force as well as the friction
coefficient for the unfolding process, and we describe the associated dynamics by
a non-linear Langevin equation. By analyzing experimental data where DNA
molecules are photo-cut and unfolded inside a nanochannel, our model allows us
to extract values for the unfolding force as well as the friction coefficient for
the first time. In order to extract numerical values for these physical
quantities, we employ a recently introduced Bayesian inference framework. We
find that the determined unfolding force is in agreement with estimates from a
simple Flory type argument. The estimated friction coefficient is in agreement
with theoretical estimates for motion of a cylinder in a channel. We further validate
the estimated friction constant by extracting this parameter from DNA's
center-of-mass motion before and after unfolding, yielding decent
agreement.  We provide publically available software for performing
  the required image and Bayesian analysis.
\end{abstract}


\maketitle

\section{Introduction}

Nanofluidic channels combined with fluorescence microscopy have, during the last decade, appeared as an experimental tool for studying the conformations of single DNA molecules under nanoconfinement{.\cite{persson2010dna,tegenfeldt2004dynamics,levy2010dna,Dorfman2013,Dai2015} This method allows to stretch the DNA molecule so that its
extension in the channel direction becomes much larger than the radius of gyration of the unconfined molecule, making it possible to obtain coarse-grained sequence information through fluorescent labeling of DNA.\cite{Jo2007,Lam2012,Mic12,Nilsson2014,Muller2017} Nanoconfined DNA has also been used to study
molecular crowding,\cite{zhang2009macromolecular,zhang2012nanouidic} and the physical properties of nanoconfined DNA-protein complexes.\cite{zhang2013nanofluidic,zhang2013effect,frykholm2014probing,persson2012lipid,roushan2014probing,Fryk16,Fryk17}

Steady-state fluctuations of DNA conformations have been investigated thoroughly during the last decade.\cite{reisner2005statics,persson2009confinement,Yang2007,burkhardt2010fluctuations,werner2012orientational,Muralidhar2014,Muralidhar2014a,Muralidhar2016,Muralidhar2016a,Wang2011,Dai2013,Dai2014,Gupta2014,Gupta2015,werner2014confined,Werner2015,Iarko2015,Reinhart2015,odman2018}
The work prior to 2012 was reviewed by \citet{Reisner2012}. The emerging picture is quite complicated. Stiffness, self avoidance, and confinement compete in determining the DNA extension, giving rise to a multitude of apparently distinct physical regimes.  Recently it was shown,\cite{Werner2017} however, that the problem of determining the steady-state extension distribution of nano-confined DNA can be mapped to a single stochastic,  telegraph model that describes the DNA conformations as random yet persistent random walks. This theory predicts the universal scaling properties of
the steady-state extension distribution over a  wide range of parameters. It is in excellent agreement with simulation results, and in good agreement with experiments. The steady-state conformation fluctuations of nanoconfined DNA molecules are thus well understood, including  solvent effects.\cite{reisner2007nanoconfinement,zhang2008effects}

Much less is known about the dynamics of confined DNA.  Just as in the steady-state case, the dynamics is very different depending on whether self-avoidance matters or not. When the DNA molecule is so strongly confined that it does not
fold back onto itself (it does not form \lq hairpins\rq{}), then the problem of determining the conformational dynamics  maps to that of a particle diffusing  in a potential \cite{Tree2013}, yielding an estimate of the relaxation time of the stochastic  extension dynamics.

In wider channels where the molecule folds back many times, by contrast, the extension dynamics was
described by  a deterministic law,\cite{levy2008entropic,ibanez-garcia_hairpin_2013} describing how the entropic force due to  self avoidance causes the molecule to unfold until the extension reaches  its steady state. The unfolding time is determined by the competition between the deterministic entropic force and hydrodynamic friction.
A related problem is the ejection of a DNA molecule from a nanochannel, also in this case the dynamics was described by a deterministic law.\cite{milchev_ejection_2010}

In general the dynamics of confined DNA is subject to diffusive fluctuations, originating from the molecular motion in the fluid. 
\citet{werner2018hairpins} considered  a single hairpin in the conformation of a confined semiflexible polymer. Entropic repulsion causes the hairpin to unfold. This process is described by a generalized diffusion equation that takes into account  the competition between the deterministic entropic  force and the stochastic molecular dynamics.\cite{werner2018hairpins}

\citet{alizadehheidari2015nanoconfined}  measured the unfolding dynamics of circular DNA to its linear form. Several different types of biological DNA molecules are circular, such as mitochondrial DNA in eukaryotic cells and chromosomal and plasmid DNA in bacteria. In particular, nanofluidic channels have been used to identify and characterize bacterial plasmids that render bacteria resistant to antibiotics.\cite{Fryk15,Mul16a,Mul16b,Nyb16} The constraint that the ends of the DNA strand must connect to form a loop reduces the steady-state extension of the DNA. When the circular molecule is cut, the resulting strand consequently extends to a larger steady-state extension.  It was observed
  that the unfolding process exhibited fluctuations that prevented a full
  quantitative analysis of the dynamics in terms of a deterministic model.
 
In this study, we developed a model to interpret the experimentally observed unfolding of circular nano-confined DNA molecules.
There are three main aspects that distinguish our system from those in Refs.~\onlinecite{Tree2013,levy2008entropic,ibanez-garcia_hairpin_2013,milchev_ejection_2010,werner2018hairpins}. First, the persistence length of the DNA molecules is of the same
order as the channel size, so that the DNA molecule can fold back upon itself several times. The resulting conformations are thus quite different from those
considered in Ref.~\onlinecite{werner2018hairpins}.
Second, immediately after the loop is cut, the unfolding process starts from both open ends. Eventually one side reaches a steady-state
extended configuration, while the other end continues to unfold. These two stages are described by different equations. Third, and most importantly, the diffusive part of the dynamics is considered, not only its deterministic counterpart.\cite{levy2008entropic,ibanez-garcia_hairpin_2013}

We compare the predictions of our model to new experimental data. The experiments reported here use the same method as \citet{alizadehheidari2015nanoconfined}, but here the DNA is much longer which makes the unfolding slower and easier to analyze in detail.
We label the DNA with the fluorescent dye  YOYO-1 and exploit the fact that this dye �forms reactive oxygen species in its excited state. These reactive molecules cause single-strand breaks\cite{tycon2012quantification,aakerman1996single} on the DNA backbone, so called \lq nicks\rq{}. When two  nicks occur sufficiently close to each other on opposite strands, a double-strand cut occurs,  that subsequently leads to unfolding of the circular DNA.  In our experiments we observe how this unfolding proceeds in real time using fluorescence microscopy. 

Using our model we construct likelihood
functions and analyze the experiments using a new Bayesian inference procedure.\cite{krog2017bayes} This framework allows us to accurately pin point the parameters in the model (unfolding force and friction constant per length of the DNA), as well as testing the quality of the stochastic model. A comparison between the inferred parameter values and simple theoretical estimates shows reasonable agreement.

\section{Experiments}

The experiments use the same method as \citet{alizadehheidari2015nanoconfined}.
Nanochannels with dimensions 100~x~150~nm$^2$ were fabricated in fused silica by following a fabrication method described elsewhere\cite{persson2010dna}. 
Circular DNA molecules of size $\sim 130$ kbp (In Ref. \onlinecite{alizadehheidari2015nanoconfined} the DNAs where 42 kbps) stained with YOYO-1 at a ratio of 1 dye per 10 bp were driven into the nanochannels by first identifying them visually in connected microchannels.
The circular DNA was unfolded to linear DNA while enclosed in the nanochannels by generating a double strand cut via irradiation with the excitation light source of the microscope.
The DNA molecules were imaged with an epifluorescence microscope using 100x oil immersion and either a Photometrics Evolve EMCCD camera (pixel size: 159.2 nm) or an Andor ixon ultra 888 EMCCD camera (pixel size: 130 nm) at a frame rate of either 9 fps or 6.2 fps with 100 ms exposure time, respectively.
Each molecule was recorded in the circular form for a few hundred of frames and then the unfolding process was
recorded and finally the linear form of the DNA was imaged until a second
double strand break occurred, see Fig. \ref{fig:experiments} for two examples.
The experiments were performed at two buffer concentrations, 0.05X TBE and 0.5X TBE which corresponds to ionic strengths 3.8 mM and 24.9 mM, respectively.\cite{Iarko2015} 
\begin{figure}[t]
\begin{center}
\includegraphics[width=0.5\textwidth]{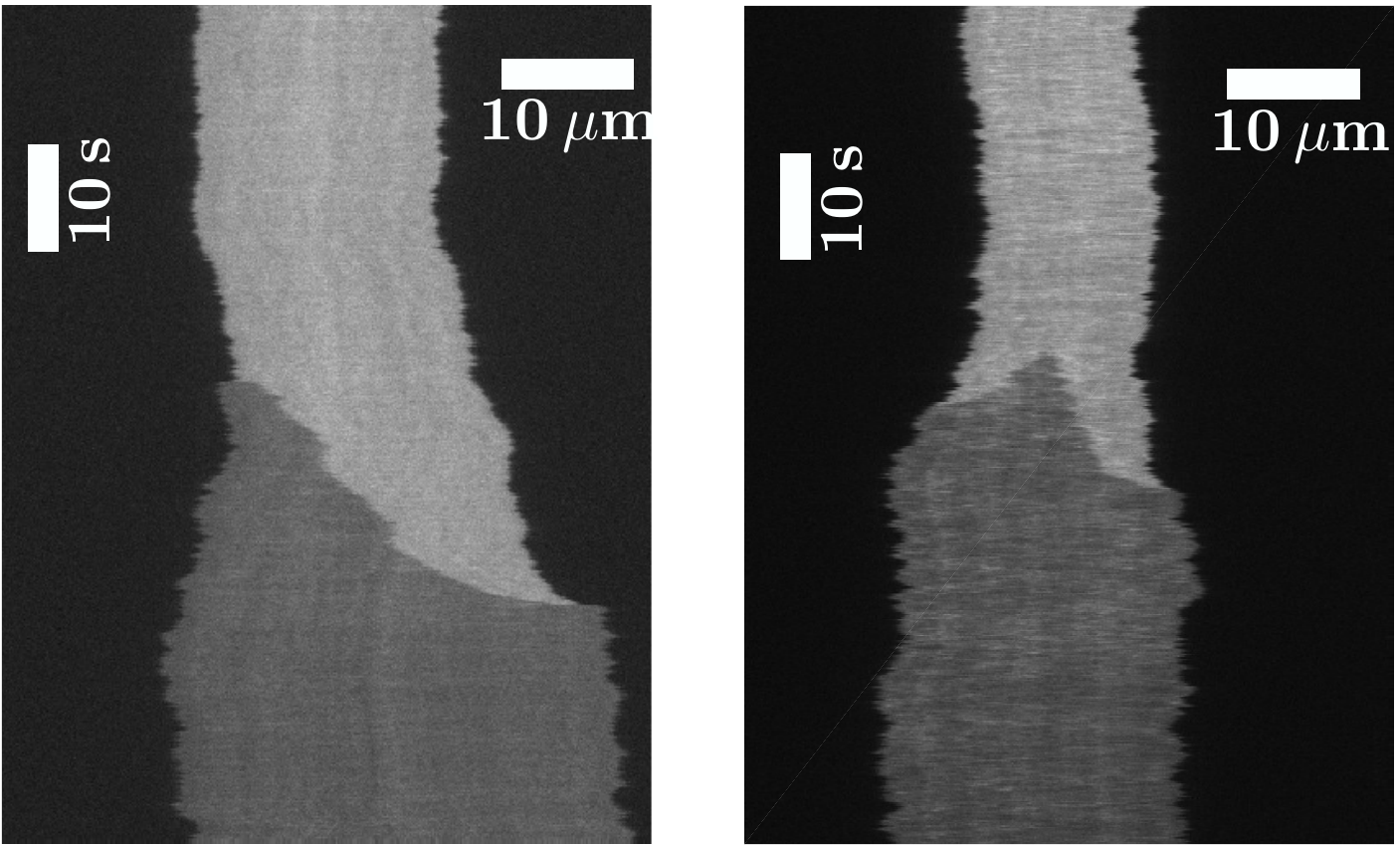}
\caption{Kymographs of plasmid unfolding in nanochannels in (left) 0.05 X TBE
  and (right) 0.5 X TBE buffer conditions (see text).}
\label{fig:experiments}
\end{center}
\end{figure}

\section{Model of the unfolding dynamics}
\label{dnamodel}

Consider a circular DNA in a nanochannel which at time $t=t_0=0$ is cut at some point along its contour. At this initial time there are two folded parts of the molecule which unfold simultaneously. The dynamics then proceeds in two stages: (i) short times where both folds are present, and (ii) later times, when only one fold is left. These feature different dynamics and are addressed individually below.

The molecule unfolds after the cut because the free energy is higher in the folded state, due to repulsive short-range interactions between the overlapping strands. Since the number of contact points is proportional to the extension of the blob, so is the extra free energy caused by overlaps. As a result, the unfolding force $f$ is independent of the extension of the blob. This unfolding force is counteracted by a drag force, caused by the movement of the unfolding DNA through the solution. Since hydrodynamic interactions are screened at length scales beyond the channel dimension \cite{muralidhar2015kirkwood,bakajin1998electrohydrodynamic},
the drag force is proportional to the extension of the unfolding part, so that its mobility $\mu$  is inversely proportional to its extension $x$, $\mu = 1/(\gamma x)$.  Here $\gamma$ is the friction coefficient.

 To obtain a tractable model of the unfolding, we make  four assumptions. First, we assume that the unfolding process is ``slow'', i.e. that it is much slower than the internal dynamics within each strand. Second, we assume that the extension of a given DNA segment is not affected by the interactions with the overlapping strand. In other words, we assume that the extension after unfolding is twice as large as  the extension of the uncut molecule, $\XU = 2X_\text{F}$ (Fig.~\ref{fig:ext}). 
 This is an oversimplification,\cite{alizadehheidari2015nanoconfined} and in
 Section~\ref{discussion} we discuss how this may affect our parameter estimates.  Third, we assume that the mechanical properties of the DNA  remain unchanged throughout the unfolding. Also this is a simplification, because the imaging of the unfolding
 DNA may cause photonicking which in turn may change the mechanical properties of the DNA molecule. Fourth, we make the assumption that during unfolding there is no friction between the "upper" and "lower" parts (see Figs. \ref{fig:ext} and \ref{fig:cartoon}) of the DNA. 

 \begin{figure}[t]
\begin{center}
\includegraphics[width = 0.48\textwidth]{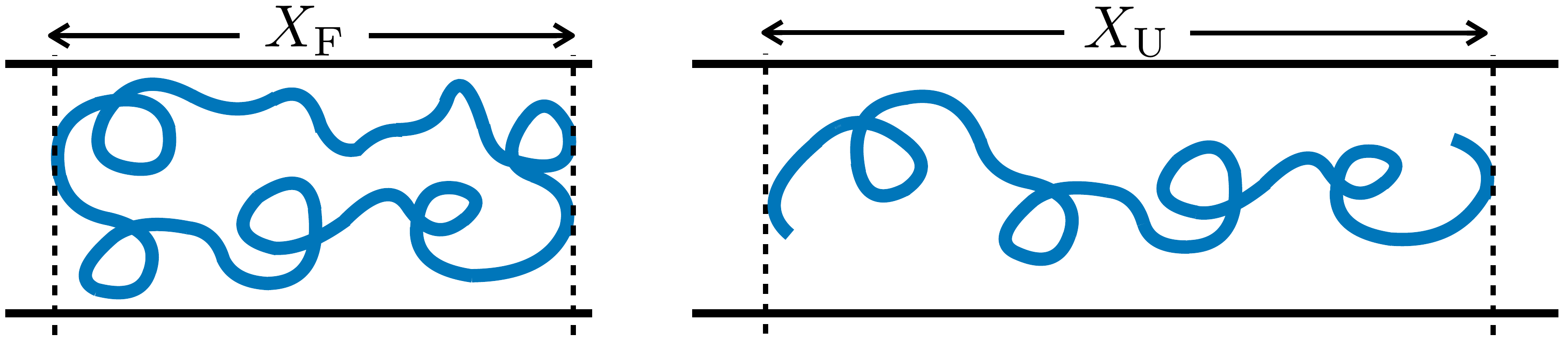}
\caption{Illustration of the extensions of circular and linear DNA in nanochannels.}\label{fig:ext}
\end{center}
\end{figure}
Note that we do not make any explicit assumptions about whether our system is in the Odijk regime\cite{Reisner2012} or in the extended de Gennes regime.\cite{Wang2011,Dai2013,Dai2014,werner2014confined} Since the channel width is of the order of the persistence length of the DNA molecule, our system is in fact in between these two regimes. However,
it was recently shown that the underlying physics is the same in the different regimes,\cite{Werner2017} 
only prefactors in the  estimates of forces and mobilities may differ. In  particular, the force $f$ is independent of the extension, and the mobility is inversely proportional to the extension. Here we determine the two parameters, $f$ and $\gamma$, directly from experiments.

\begin{figure}[h]
\begin{center}
\includegraphics[width = 0.4\textwidth]{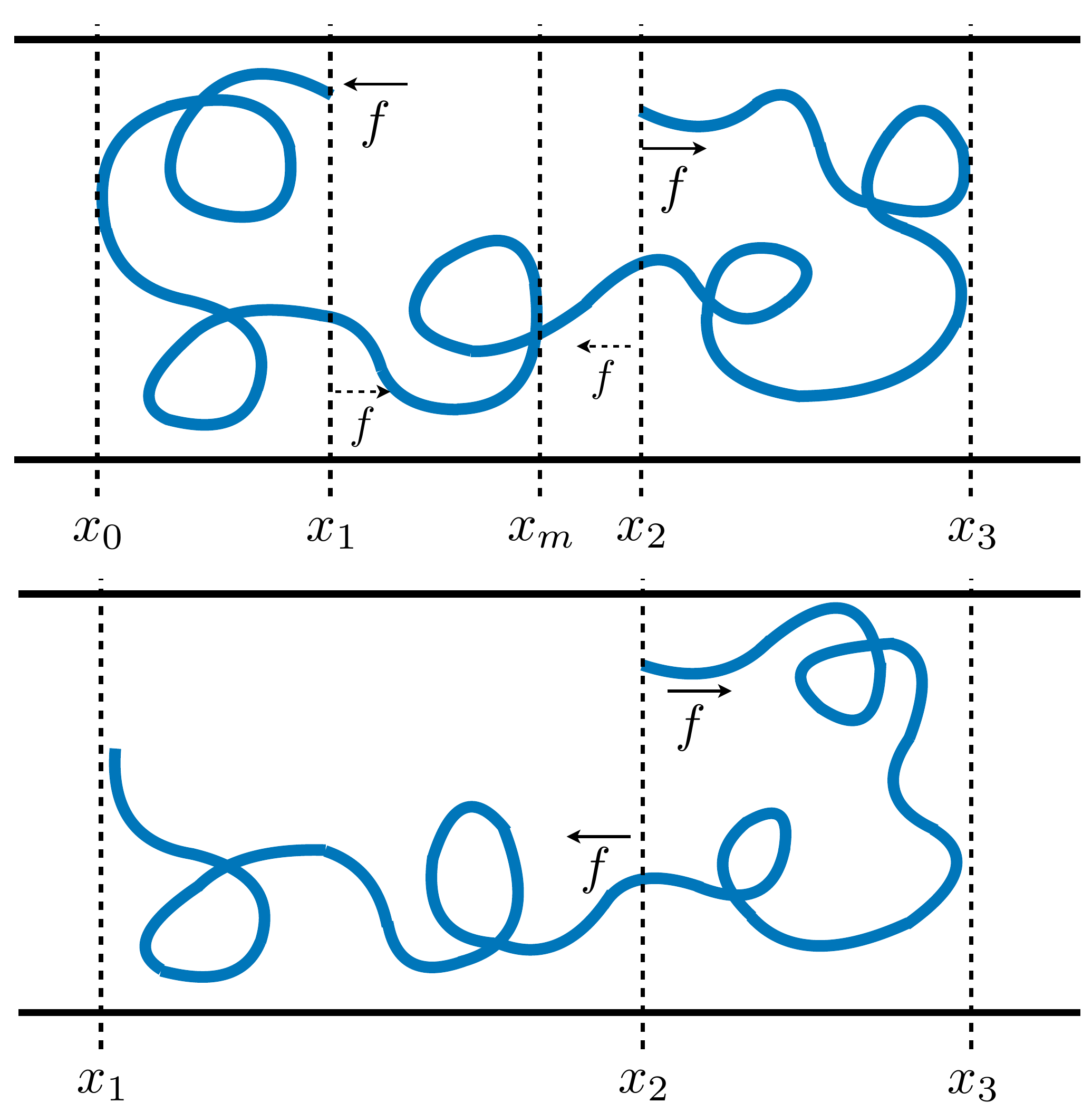}
\caption{Cartoon of DNA unfolding. A circular DNA molecule is cut at a random
  point along its contour. There are unfolding forces (all with magnitude $f$) of entropic
  origin that drives the process, counteracted by frictional forces. The
  dynamics proceeds in two stages: In stage 1 (top) the DNA consists of two
  folded parts, while in stage 2 (bottom) one of these has unfolded. The
  unfolding proceeds until the conformation is linear.  }
\label{fig:cartoon}
\end{center}
\end{figure}
\subsection{Double fold (stage 1)}
We characterize the unfolding  of the DNA blob by the coordinates
$x_0,x_1,x_2,x_3,x_m$, Fig.~\ref{fig:cartoon} (top).
These coordinates are time-dependent, but we
 leave the time-dependence implicit in this section, for convenience.
We define the coordinate vector
\begin{equation}
\mathbf{x} =
\begin{pmatrix}
x_1 \\ x_2 \\ x_m
\end{pmatrix}\,,
\end{equation}
where $x_1$ and $x_2$ are shown in Fig.~\ref{fig:cartoon} (top), and $x_m$ denotes the coordinate of the center of mass of the DNA blob
(it will be integrated out later). The associated forces are
\begin{equation}
\label{uforces}
\vf =
\begin{pmatrix}
-f \\ f \\ 0
\end{pmatrix},
\end{equation}
The first component of $\vf$ is the force $f$ acting upon the upper left part of the DNA in Fig.~\ref{fig:cartoon} (top). The second component of $\vf$ is the force acting upon the upper right part of the DNA. Notice that these two forces have different signs, but  the same magnitude (since they originate from the same physical mechanism, see appendix \ref{estimates:force}).  The 
force associated with the center-of-mass coordinate is the sum of all forces acting the different parts of the DNA, see Fig.~\ref{fig:cartoon} (top), and hence the third component in $\vf$ vanishes (there is no net force acting on the DNA).  Also notice the dashed force arrows in Fig.~\ref{fig:cartoon} (top). These cancel out and, for a system where the polymer extension is independent of the configuration (such that $\XU = 2X_\text{F}$), they do not contribute to the dynamics in our model.

Given our choice of coordinates and associated forces, the system obeys the Fokker-Planck equation\cite{van1992stochastic}
\begin{equation}
\partial_t p(\pmb{x},t) = -\pmb{\nabla}^T(\pmb{\mu}\,\vf\, p - k_BT\pmb{\mu}\pmb{\nabla} p).
\end{equation}
The mobility matrix, $\pmb{\mu}$, is defined through
\begin{equation}\label{eq:mob_matrix_def}
\langle \dot{\mathbf{x}} \rangle = \pmb{\mu}\vf,
\end{equation}
 where $\langle\ldots\rangle$ denotes ensemble average, i.e., an average over different realizations of the process with identical initial conditions.  
 
Let us now consider the mobility matrix. 
As we focus on the unfolding dynamics, we consider the lengths of the folded
parts of the molecule (Fig. \ref{fig:cartoon}), i.e. the \textit{folds}, which have extensions
\begin{equation}
\label{xlxr}
x_l = x_1 - x_0\,, \quad x_r = x_3 - x_2,
\end{equation}
corresponding to the left and right fold, respectively.
Assuming no friction or tension between the strands, the tensor $\pmb{\mu}$ is diagonal:
\begin{equation}
\pmb{\mu} = 
\begin{pmatrix}
\mu_l & 0 & 0 \\
0 & \mu_r & 0 \\
0 & 0 & \mu_m
\end{pmatrix}.
\end{equation}
Since the mobility of a segment is inversely proportional to its extension,  we must have
\begin{equation}
\mu_l = \frac{1}{\gamma x_l } \,, \quad \mu_r = \frac{1}{\gamma x_r }\,
\end{equation}
Here, $\mu_l$ and $\mu_r$ are the mobilities of the two upper segments and $\mu_m = 1/(\gamma X)$ is the mobility of the lower segment, where $X$� is the length of this segment (equal to the extension of the DNA, see Fig.  \ref{fig:cartoon}, top). We must have 
$\XU = X + x_l + x_r$ and hence 
\begin{equation}
\mu_m = \frac{1}{\gamma (\XU-x_l-x_r) }.
\end{equation}
This concludes our discussion of the mobility matrix.

Using Eq.~(\ref{Lest})} we can express the extensions of the different parts in Fig.~\ref{fig:cartoon} as
\begin{equation}
\mathbf{x}' = 
\begin{pmatrix}
x_l \\ x_r \\ x_m
\end{pmatrix}
= \mathbf{M} \mathbf{x} +
\begin{pmatrix}
 \frac{\XU}{4} \\ \frac{\XU}{4} \\
0
\end{pmatrix}
\quad \text{with} \quad \mathbf{M} =
\frac{1}{2}
\begin{pmatrix}
1 & 0 & -1 \\
0 & -1 & 1 \\
0 & 0 & 2 \footnotemark
\end{pmatrix},
\addtocounter{footnote}{-1}
\footnotetext{This value was incorrectly given as 1 in the initial arXiv
  version and in J. Krog et al.,  J. Chem. Phys. 149, 215101 (2018). This typo does not affect
  any of the subsequent equations or results. }
\end{equation}
\begin{equation}
\pmb{\mu}' = \mathbf{M}\pmb{\mu}\mathbf{ M}^T =\frac{1}{4} 
\begin{pmatrix}
\mu_l + \mu_m & -\mu_m & -2\mu_m \\
-\mu_m & \mu_r + \mu_m & 2\mu_m \\
-2\mu_m & 2\mu_m & 4\mu_m
\end{pmatrix}.
\end{equation}
In this basis,
the Fokker-Planck equation takes the form
\begin{equation}
\label{newfpe}
\partial_t p = -\pmb{\nabla}'^T(\pmb{\mu}' \vf' p - k_BT\pmb{\mu}'\pmb{\nabla}' p), \quad \vf' = \begin{pmatrix}-2f \\-2f \\0 \end{pmatrix}.
\end{equation}
Eq.~\ref{newfpe} is a generalization of Eq.~12 in
Ref.~\onlinecite{werner2018hairpins} in two ways. First, we describe the
dynamics of a system with two folds instead of one, and second, we do not
restrict the movement of the long strand of DNA connecting the folds.

Since $\pmb{\mu}'$ does not depend on $x_m$, we can integrate the Fokker-Planck equation~\eqref{newfpe} with respect to $x_m$. Removing surface terms we find for $\pmb{x}'_2 = \begin{pmatrix} x_l & x_r \end{pmatrix}^T$;
\begin{equation}
\partial_t p(\pmb{x}'_2,t) = -{\nabla'_i} \left(\mu_{ij}f'_j p(\pmb{x}'_2) - k_BT\mu'_{ij}\nabla'_j p(\pmb{x}'_2)\right), 
\end{equation}
where $i=1,2$, and we sum over repeated indices.
From this we obtain the Langevin equation for the dynamics\cite{gardiner}
\begin{equation}
\label{leeq_stage1}
dx'_i = \left(\mu_{ij}F'_j + k_BT\nabla'_j\mu'_{ij}\right)dt + \sqrt{2k_BT} B_{ij}d\eta_j,
\end{equation}
with $(\pmb{B}\pmb{B}^T)_{ij} = \mu_{ij}'$ and $d\pmb{\eta}$ is uncorrelated
Gaussian noise{, such that $\langle d\eta_i(t)d\eta_j(t') \rangle = \delta_{ij}\delta(t-t')dt$.}  
In Eq.~\eqref{leeq_stage1} the two first terms on the right hand side contribute to a mean decrease of the fold lengths $x_i$, mainly due to the unfolding force $f$. The final term on the right hand side represents the random fluctuations present in the system. The mean behavior given by the first terms depend explicitly on the current coordinates, which feature random fluctuations. 

For the analysis in Sec. \ref{bayes_est} we need the \textit{likelihood}
function  for the parameters $\gamma$ and $f$ for stage 1. In order to
calculate this quantity, we note for a small $\Dt$, we can approximate the average behavior
\begin{equation}
\label{meansoflr}
\langle x'_i(t+\Dt) \rangle = x'_i(t) + \left(\mu'_{ij}(t)F'_j + k_BT\nabla_j\mu'_{ij}(t)\right)\Dt, 
\end{equation}
where we indicate that the mobilities are calculated prior to the
increment. By defining $\pmb{y}(t+\Dt) = \pmb{x'}(t + \Dt) - \langle
\pmb{x}(t+\Dt) \rangle$, we find the covariance
\begin{equation}
\label{yvars}
\langle y_i(t +\Dt)y_j(t+\Dt) \rangle = 2k_BT\mu'_{ij}(t)\Dt.
\end{equation}
The residuals $\pmb{y}$ from a time series of $x_l$ and $x_r$, obtained via Eq.~\eqref{meansoflr}, thus
 occur with the probability\footnote{The square root in this expression was
   incorrectly shown cover the entire first factor in the initial arXiv
   version and in J. Krog et al., J. Chem. Phys. 149, 215101 (2018). This typo does not affect
  any of the subsequent equations or results.}
\begin{equation}
\label{likestage1}
p(\pmb{y}|\gamma,f) = \frac{1}{4\pi k_B T\Dt\sqrt{ |\pmb{\mu}'_2|}}
\text{exp}\left(-\frac{1}{4k_BT\Dt}\pmb{y}^T{\pmb{\mu}'_2}^{-1} \pmb{y} \right),
\end{equation}
where $\pmb{\mu}'_2$ is the upper left 2x2 matrix in $\pmb{\mu}'$. Since the thermal noise generating the deviations from the expected unfolding behavior is uncorrelated in time, the likelihood for the entire first unfolding stage is the product of the likelihoods in the form of Eq.~\eqref{likestage1}.

\subsection{One fold (stage 2)} 

For the single fold situation we define our coordinate system as  (see Fig.~\ref{fig:cartoon}, bottom)
\begin{equation}
\mathbf{x} =
\begin{pmatrix}
x_1 \\ x_2
\end{pmatrix}
\end{equation}
 with associated forces 
\begin{equation}
\label{uforces2}
\vf=
\begin{pmatrix}
-f \\ f
\end{pmatrix}
\end{equation}
 The first component of $\mathbf{f}$ above is the force acting on the upper part of the DNA in Fig.~\ref{fig:cartoon} (bottom). The second component of  $\mathbf{f}$ is the force acting on the lower part of the DNA.

The $\pmb{\mu}$ is the mobility matrix is defined through Eq. (\ref{eq:mob_matrix_def}).
Assuming, as before, that there is no friction between the strands, 
the mobility matrix is diagonal:
\begin{equation}
\pmb{\mu} = 
\begin{pmatrix}
\mu_1 & 0 \\
0 & \mu_2
\end{pmatrix},
\end{equation}
where 
\begin{equation}
 \mu_2 = \frac{1}{\gamma x }\,. 
\end{equation}
Here, $\mu_2$ is the mobility of the upper part of the DNA in Fig. \ref{fig:cartoon} (bottom). Similarly the mobility for the lower DNA part is $\mu_1 = 1/(\gamma X)$, where $X$� is the length of this segment.  Since $\XU = X+x  $  (assuming that, as before, that the extension of a DNA segment is approximately
independent on whether it is folded or unfolded) we have
\begin{equation}
\mu_1 = \frac{1}{\gamma(\XU-x) }.
\end{equation}
Given the coordinates, forces and mobilities above, the Fokker-Planck
equation for stage 2 becomes
\begin{equation}
\partial_t p = -\pmb{\nabla}^T(\pmb{\mu} \vf p - k_BT\pmb{\mu}\pmb{\nabla} p).
\end{equation}
In terms of $x_1$ and $x_2$, the measured extension of the fold is given by:
\begin{equation}
\label{x_stage2}
x = x_3 - x_2 
= \frac{1}{2}(\XU + x_1 - x_2),
\end{equation}
We define a second coordinate $z=\frac{1}{2}(x_1 + x_2)$, such that
\begin{align}
\mathbf{x}'& = 
\begin{pmatrix}
x \\ z
\end{pmatrix} 
= \mathbf{M} \mathbf{x} +
\begin{pmatrix}
 \frac{\XU}{2} \\
0
\end{pmatrix}
\quad \text{with} \quad \mathbf{M} =
\frac{1}{2}
\begin{pmatrix}
1 & -1 \\
1 & 1
\end{pmatrix},\\
\vf' &= 
\begin{pmatrix}
-2f \\ 0
\end{pmatrix},\\
\pmb{\mu}' &= \mathbf{M}\pmb{\mu}\mathbf{ M}^T =\frac{1}{4} 
\begin{pmatrix}
\mu_1 + \mu_2 & \mu_1 - \mu_2 \\
\mu_1 - \mu_2 & \mu_1 + \mu_2
\end{pmatrix} \equiv
\frac{1}{4}
\begin{pmatrix}
\mu_+ & \mu_- \\
\mu_- & \mu_+ 
\end{pmatrix}
.
\end{align}
Note that $\mu'$ is independent of the center of mass coordinate $z$.
Integrating the Fokker-Planck equation
\begin{equation}
\label{newfpestage2}
\partial_t p = -\pmb{\nabla}'^T(\pmb{\mu}' \vf' p - k_BT\pmb{\mu}'\pmb{\nabla}' p).
\end{equation}
 over $z$ and removing surface terms we find
\begin{align}
\label{fpnoab}
\partial_t p(x,t)  = 
 - \partial_x  \left( \frac{-2f}{4}\mu_+p(x,t) - \frac{k_BT}{4}\mu_+\partial_x  p(x,t) \right),
\end{align}
{which is a generalization of Eq. 12 in Ref. \onlinecite{werner2018hairpins}, since we allow movement of both strands in the fold.}
This corresponds to the Langevin equation
\begin{equation}\label{eq:le_stage2}
dx =\left(-\frac{2f}{4}\mu_+ + \frac{k_BT}{4}\partial_x\mu_+\right)dt + \sqrt{\frac{k_BT\mu_+}{2}}d\eta(t),
\end{equation}
{where $\langle d\eta(t) \rangle = 0$ and $\langle d\eta(t)d\eta(t')\rangle = \delta(t-t')dt $.}
As in stage 1, the first two terms on the left hand side of Eq.~\eqref{eq:le_stage2} are deterministic contributions to the dynamics, while the last term represents the thermal fluctuations which play a key role in the dynamics. Once again, we are forced to calculate short term expectations and variances, and cannot neglect the noise term.

In order to obtain the likelihood function we proceed as for stage 1. We use that for a short time $\Dt$ we can approximate the mean
\begin{equation}
\label{stage2mean}
\langle x(t+\Dt)\rangle = x(t) + \left( -\frac{2f}{4}\mu_+ + \frac{k_BT}{4}\partial_x\mu_+\right)\Dt 
\end{equation}
and by defining $y(t+\Dt) = x(t + \Dt) - \langle x(t +\Dt) \rangle$, we find
\begin{equation}
\label{stage2var}
\langle y(t+\Dt)^2 \rangle = \frac{k_BT}{2}\mu_+\Dt,
\end{equation}
such that the likelihood function for a specific increment
\begin{equation}
\label{likestage2}
p(y|\gamma,{f}) = \frac{1}{\sqrt{\pi k_BT\Dt \mu_+}}\text{exp}\left(-\frac{y^2}{k_B T\Dt\mu_+}\right),
\end{equation}
where $\mu_+$ in \eqref{stage2mean} is, once again, evaluated at the coordinate before the increment.

Since each individual increment is taken to be independent, the complete likelihood function is the product of the likelihoods of stage 1 and stage 2.

\section{Bayesian parameter estimation}
\label{bayes_est}

\begin{figure}
\center
\includegraphics[width=0.5\textwidth]{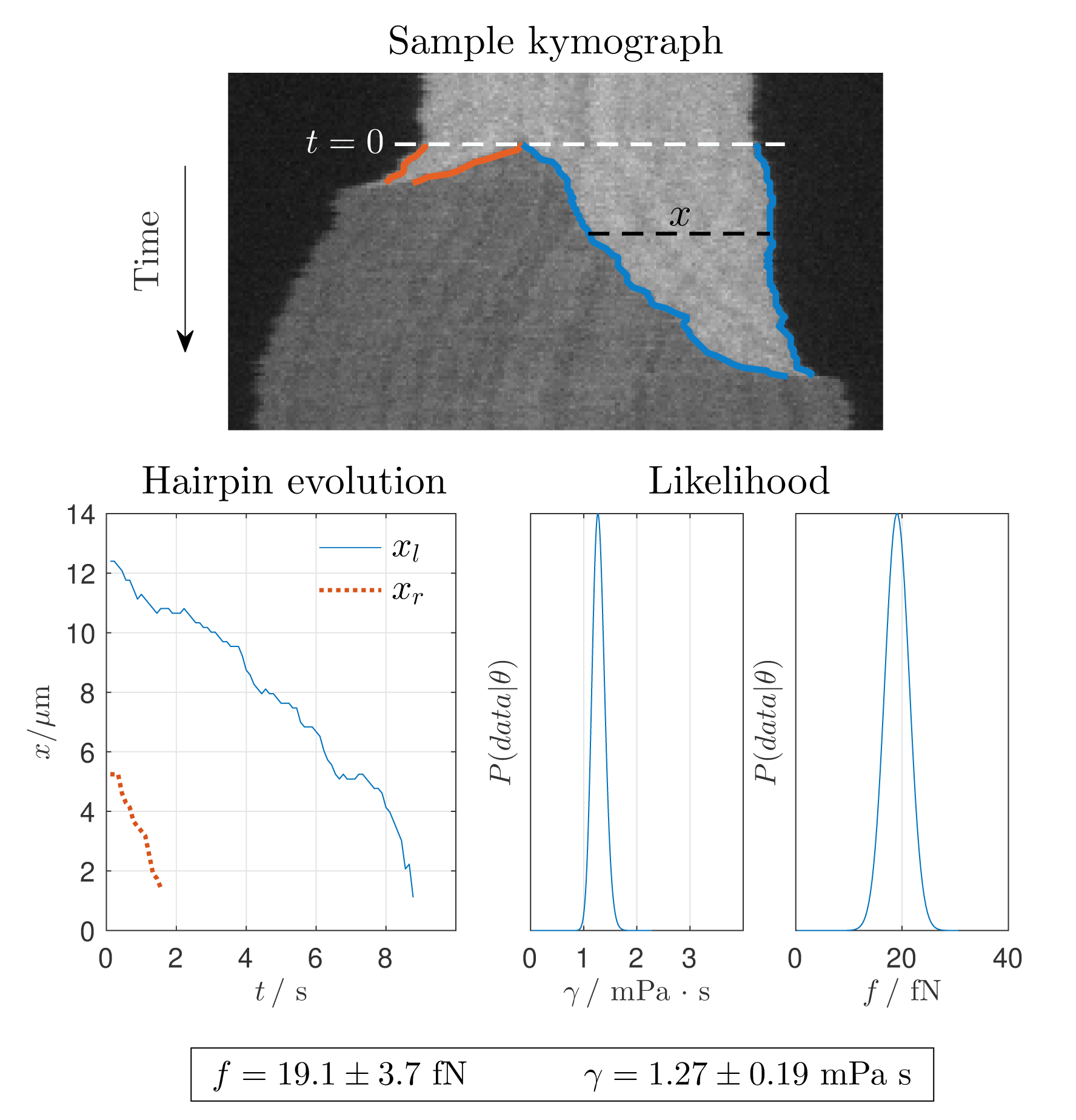}
\caption{(Top) Sample kymograph with the extracted fold contours
  highlighted. The columns in this image correspond to different
  positions along the nanochannel, and the rows corresponds to different time
  frames. Dark regions are background, intermediate intensities correspond to linear
  DNA, and high intensity regions are double-stranded DNA.
 (Bottom, left) Time series of the fold lengths during the unfolding
 process. (Bottom, right) Slices of the likelihood functions for the friction
 coefficient, $\gamma$, and unfolding force, $f$ around the maximum likelihood point, obtained using Bayesian
 inference on the hairpin trajectories to the left.}
\label{samplelike}
\end{figure}

Armed with a stochastic model for the unfolding process, we utilize the Bayesian framework to infer parameter estimates for
experimental data. For each individual experiment we extract the coordinates
$x_0(t),x_1(t),x_2(t),x_3(t)$ at each measurement, as seen in Fig.~\ref{samplelike}. We then use Eq.~\eqref{xlxr} to find $x_l(t)$
and $x_r(t)$ in stage 1  and Eq.~\eqref{x_stage2} to find $x(t)$ in stage 2. 
 In order to estimate $\XU$, we
take the time average of the observable 
\begin{equation}
\label{Lest}
\tilde{X}_U(t) = 2x_3 - x_2 + x_1 - 2x_0
\end{equation}
for the stage 1 dynamics. We, however, point out that there are several potential estimates of $\XU$. For instance, we could take the time average of the extensions after complete unfolding, or, twice the time average of the measured extension before the cut occurred. These two estimates yield slightly different results. The estimate above is a compromise between these two cases. In Sec. \ref{discussion} we comment on how the assumption that $\XU$ remains constant during unfolding may affect estimated parameters. 

To infer the parameters $\pmb{\theta} = (f,\gamma)$ from the
data $D = \{x_l(t),x_r(t),x(t)\}$ $\forall t$, we employ Bayes' theorem
\begin{equation}
\label{posterior}
P(\pmb{\theta}| D ) = \frac{P(D|\pmb{\theta})P(\pmb{\theta})}{P(D)},
\end{equation}
where $P(D|\pmb{\theta})$ is the \textit{likelihood function} and $P(\pmb{\theta})$ is the prior probabilities for the parameters. The left hand side is the posterior probability distribution for the parameters and describes our knowledge of the parameters knowing the experimental data.\cite{gelman2014bayesian}
For the sake of parameter estimation, we may consider the \textit{evidence}
$P(D)$ as being a normalization constant, whereas for model comparison it is
the main subject of interest.

If the distribution in Eq.~\ref{posterior} has a single peak, then its mean constitutes our best guess for the true parameters $f$ and $\gamma$, while the peak width determines the uncertainty of the estimate.
As long as the prior $P(\pmb{\theta})$ is nonzero around the peak of the
likelihood function, the details of its shape has little influence on the
results. We choose a uniform prior for the unfolding force such that $f \in
[0;0.500]$ pN\footnote{The unit was incorrectly given as fN in the initial
  arXiv version and in J. Krog et al., J. Chem. Phys. 149, 215101 (2018).}, while we use a Jeffreys prior\cite{gelman2014bayesian} for the
friction and restrict it such that $\gamma \in [10^{-6};0.5]\,
\text{Pa}\cdot\text{s}$. The Jeffreys prior assigns equal prior probability to each decade, and is used since the order of magnitude for the parameter is unknown.

The shape of the posterior distribution in Eq.~\eqref{posterior} is found by calculating the likelihood function for each individual kymograph. If $T$ timesteps $\Delta D$ are extracted from the kymograph, where $T_2$ of them show two folds, then the total likelihood function is given
\begin{equation}
\label{totall}
P(D|\pmb{\theta}) = \prod_{i=1}^{T_2}P(\Delta D_i|\pmb{\theta})\prod_{j=T_2+1}^{T}P(\Delta D_j|\pmb{\theta}),
\end{equation}
where the $P(\Delta D_i)$ in the first sum is calculated with Eq.~\eqref{likestage1} and $P(\Delta D_j)$ in the second sum is calculated with Eq.~\eqref{likestage2}.
We use the Nested Sampling algorithm\cite{skilling2006nested} to investigate the shape of the posterior Eq.~\eqref{posterior} which is done by exploring the likelihood Eq.~\eqref{totall} for all regions of the parameter space allowed by the prior and increasing sampling density as the likelihood rises around the peak.
We display a sample data set along with slices of its likelihood function in
Fig.~\ref{samplelike}, where the parameters are varied around the maximum likelihood point to display how the likelihood decays in both directions.
The likelihood functions shown above should be interpreted with care, since they are one dimensional slices of a two-dimensional parameter space, but they serve as a useful visual guide. Printed estimates reflect the complete set of samples.

In addition to parameter estimation, we perform a model check to confirm the validity of the model for the dynamics. This is done by drawing sets of parameters $\pmb{\theta}^*$ from the inferred posterior distribution in Eq.~\eqref{posterior} and simulating a new data series $D^*$ from the same initial coordinates as the original data.
We compare the new likelihood $P(D^*|\pmb{\theta}^*)$ with the original $P(D|\pmb{\theta}^*)$ and extract p-values as described in Ref. \onlinecite{krog2017bayes}.
If the model is a good description of the data, then the data set $D$ and its
corresponding likelihood should be \textit{typical} for the model, and we expect that
the p-value is not close to either 0 or 1. In contrast if the model lacks a
critical aspect of the dynamics, then extreme values of $p$ are expected to
occur. 

Matlab-based image analysis software for edge tracing in
  experimental  kymographs  (see appendix
  \ref{image_analysis}) and software (Matlab) for performing  the subsequent
  Bayesian analysis is available at {\tt https://github.com/krogjens/dna\_unfolding}. 

\section{Results}

In total 11 time traces of unfolding DNA for salt concentration 0.05X TBE and 9
time traces for the concentration of 0.5X  TBE were collected and analyzed. 
The kymographs
(compare to Figure \ref{samplelike}) were pre-processed, as shown in
appendix~\ref{image_analysis} to generate time series for the fold lengths discussed in section~\ref{dnamodel}. 

\begin{figure}[h]
\center
\includegraphics[width=0.45\textwidth]{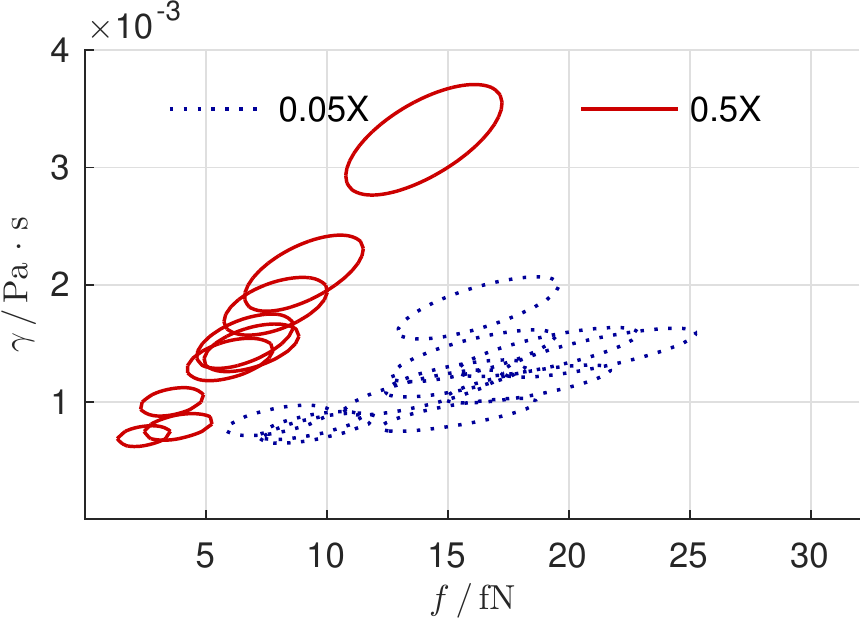}
\caption{Likelihood contours, as in Eq.~\eqref{totall}, for the individual data sets for two salt concentrations drawn at $\text{exp}(-\frac{1}{2})$ times the maximum value.}
\label{circles}
\end{figure}

The likelihood for each individual data set was explored for the parameter space as specified in section~\ref{bayes_est}.
In Fig.~\ref{circles} we display contours of the likelihood function where the contour is drawn at $\text{exp}(-\frac{1}{2})$ of the peak value. Assuming a Gaussian shape of the likelihood, each contour corresponds to the 1$\sigma$ curve and holds approximately $39.4\%$ of the likelihood for the data set.

Visual inspection indicates that data sets from different salt concentrations follow two different distributions while being stretched along a line through the origin due to some systematic effect. We have checked that similar spreads occur for sets of artificial data using fixed values for $\gamma$ and ${f}$, and illustrate this in Appendix~\ref{sims}.
 
For a compiled result we utilize the mean values for the parameters $f$ and $\gamma$ from the Bayesian analysis of each data set. We then average over the data sets for each concentration and display the results in Table~\ref{meansandvars}. Since only the ratio $f/\gamma$ has been estimated in previous experiments, we add such estimates to the table.
\begin{table}[h]
\begin{center}
\begin{tabular}{|c|c|c|c|}
\hline
Ionic strength & $f / \text{fN} $ & $\gamma / \text{mPa s}$ & $\frac{f}{\gamma} \,/\, \frac{\mu\text{m}^2}{s}$ \\
\hline
0.05X & $14.8\pm 1.3$ &$ 1.16\pm 0.10$ &$12.8 \pm 0.8$  \\
\hline
0.5X & $6.7\pm 1.2$ &$1.56\pm0.26$ &$ 4.25 \pm 0.16$   \\
\hline
\end{tabular}
\caption{The mean values and standard errors for the unfolding force and friction coefficient for each salt concentration estimated from the unfolding dynamics.
}
\label{meansandvars}
\end{center}
\end{table}

Let us now compare our results Table \ref{meansandvars} to theoretical predictions for the unfolding force and friction constant. In appendix \ref{estimates:force}, a Flory-type argument yields an estimated unfolding force in the extended de Gennes regime, see Eq. \eqref{f_est}. Using this expression, we find the estimates  $f^{0.05X} = 9.4 \text{ fN}$ for salt concentration 0.05X TBE and $f^{0.5X} = 6.2 \text{ fN}$
 for salt concentration 0.5X TBE. These values are somewhat  smaller than the experimentally inferred ones, while still within an order of magnitude.
Similarly we can provide a simple theoretical estimate for the friction. The estimated drag for a simplified steady state model of cylinders shown in appendix~\ref{estimates:gamma}, Eq. \eqref{gam_est}, displaying only a weak logarithmic dependence on the effective width and hence ionic strength. Plugging in numerical values for the quantities in this equation, we obtain $ \gamma^{est} \simeq 1.6 \text{  mPa s}$.  
The Bayesian estimates are thus in reasonable agreement with naive estimates, while also indicating the anticipated effect of changing salt concentrations.

In order to validate the estimated numerical value for $\gamma$ above, we
estimated $\gamma$ in two other independent, and simpler, ways. 
To that end, for the experiments where the edge coordinates of the molecule could be
extracted at all times, we carried out the the Bayesian data analysis on
the Brownian motion of the center of mass coordinate before and after the
unfolding takes place (motion of circular DNA and fully unfolded linear DNA). From this the total friction constant for the DNA could
be obtained, and by dividing this number by $\XU$, we find two other estimates for
the friction coefficient, $\gamma$. By this approach, we find values of
$\gamma$ that are higher than the ones obtained from the unfolding analysis, see Table.~\ref{comtab}.
\begin{table}[h]
\begin{center}
\begin{tabular}{|c|c|c|}
\hline
Ionic strength &  \,$\gamma_{\text{pre cut}} / \text{mPa s}$ & \,$ \gamma_{\text{post cut}} /\text{mPa s} $ \\
\hline
0.05X & $6.03 \pm 0.32$ & $3.03\pm 0.20$ \\
\hline
0.5X &$8.93\pm 0.21$ &$ 3.19\pm0.08$   \\
\hline
\end{tabular}
\caption{The mean values and standard errors for the friction coefficient for each salt concentration estimated from the center of mass dynamics.
}
\label{comtab}
\end{center}
\end{table}
The discrepancy between the friction coefficient estimates for the different
regimes is concerning, but, we believe, understandable. We address possible
causes in Section~\ref{discussion}.

\begin{figure}[]
\center
\includegraphics[width=0.45\textwidth]{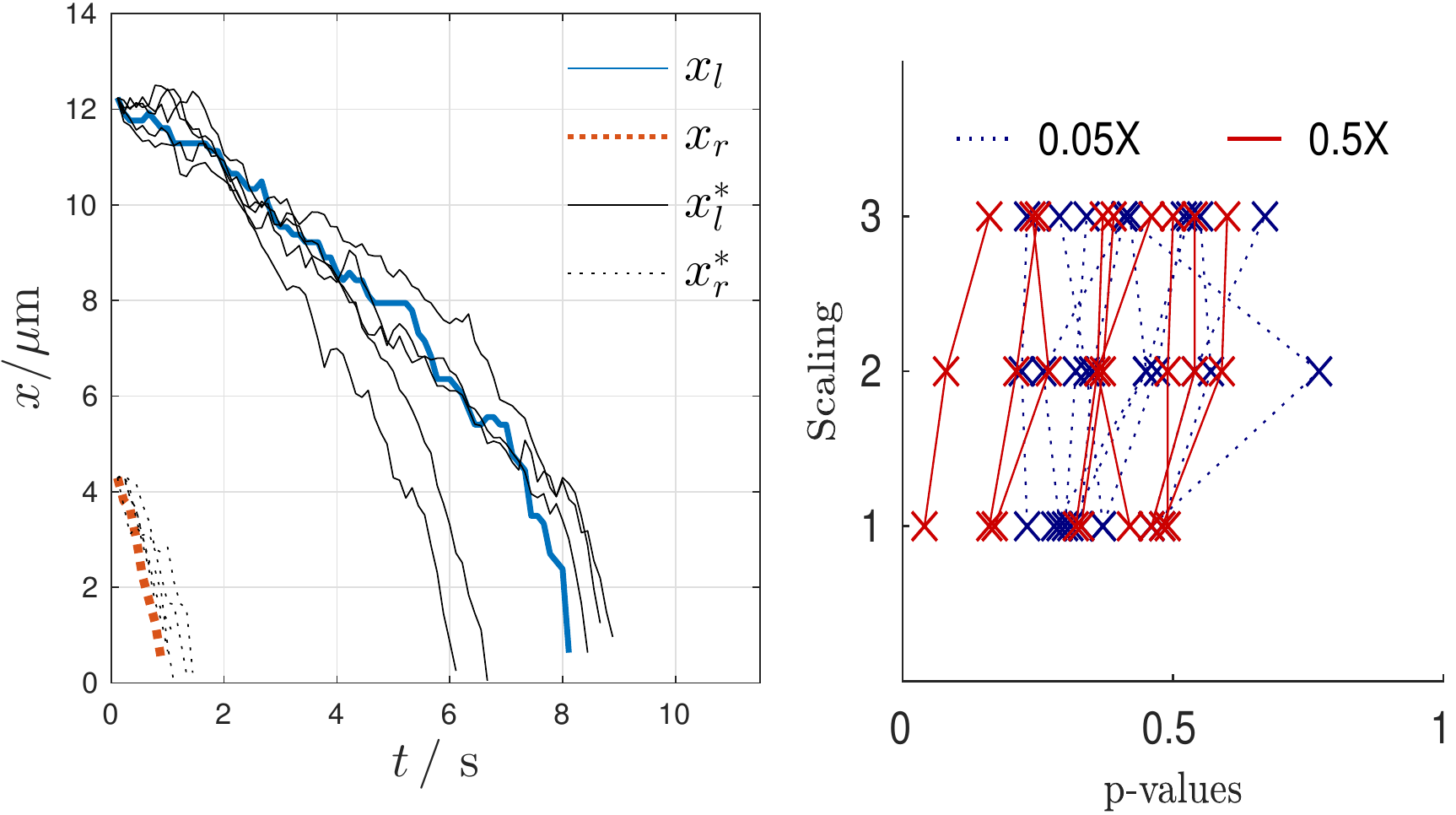}
\caption{(Left) An experimental sample trajectory along with simulated trajectories using the same starting point and the  estimated values for $f$ and $\gamma$ in the model of Section.~\ref{dnamodel}.
(Right) p-values for this model  are shown for each data set with rescaling two and three times the time step.}
\label{pvals}
\end{figure}

Performing the information content check described above on each data set, we obtain the p-values in Fig.~\ref{pvals}, depicting the original data set as well as two rescalings.
As expected for a model which describes the data well, the p-values seem centered around 0.5 with few extreme values for the model.
Note that nearly all p-values increase at the first rescaling step, in such a way as to centralize the distribution somewhat. This could be an indicator of some short time dynamics, e.g. measurement noise, which we cannot account for.

\section{Summary and discussion}
\label{discussion}
We introduced a stochastic model for the unfolding of polymers in narrow
channels. Using the Bayesian
framework we have used the model to perform parameter inference for
experimental data. Our framework allows for inference of the friction coefficient as
well as the unfolding force, which have not previously been obtained
independently. Comparing the experimental results to theoretical estimates, we
find that the parameters show reasonable agreement. 

To validate the results, we also estimated the friction coefficient from the
regular {center-of-mass} diffusion of the DNA molecules prior to and after the unfolding
process. We found consistently larger values for
the friction coefficient, while remaining within one order of
magnitude. Several factors may produce this discrepancy. For instance, in the
image analysis, the trajectories are extracted by identifying areas with
different light intensity. The lack of contrast between the circular DNA,
linear DNA and background regions can cause error and associated
``jaggedness'' in the estimated edge positions in the kymographs. Such
``jaggedness'' will appear as effective noise in the model, and in effect lead
to a reduction in the estimated friction coefficient.
We also neglect DNA-wall
  interactions in our model. A circular DNA molecule is more
  likely to have DNA segments close to the channel wall. Therefore, DNA-wall
  interactions could be more pronounced for circular DNA compared to linear
  DNA. This could be possibly, in part, explain the larger friction
  coefficient for intact circular DNA compared to completely linearized DNA (See Table \ref{comtab}).

Another shortcoming of our model is that,  as in the study by Levy et al.\cite{levy2008entropic},  we assume that the extension of a DNA segment is unaffected by interactions with other overlapping strands.
  This assumption is, however, not correct in
general, since entropy effects  cause the local extension of a segment to expand when there are other strands nearby. Thus, if the DNA has contour length $L$ and is in a linear (unfolded) state in
the nanochannel, the extension $X_{\text{U}}$ is given as $X_{\text{U}} =
\alpha_{\text{U}}L$. In the extended de Gennes regime, the value of the prefactor $\alpha_{\text{U}}$
is known exactly,\cite{werner2014confined} and for stronger confinement the values were computed in
Ref.~\onlinecite{Werner2017}. In the circular (folded) state, the molecule has extension 
$ X_{\text{F}}=\alpha_{\text{F}}\frac{L}{2}$,
see Fig.~\ref{fig:ext}.
 Dorfman {\em et al.} \cite{dorfman2018odijk} report that for their measurements, $\alpha_{\text{U}}$~is between 0.83 and 0.90, while estimates of
the ratio $\alpha_{F}/\alpha_{U}$ range between $1.36\pm0.14$ for 183 nm channels
and $1.12 \pm 0.26$ for 45 nm channels. 
We also estimated the extension ratios for out data (same trajectories which were used in the center-of-mass analysis, see appendix \ref{sec:com_analysis}). We found that for  0.05X TBE  $\alpha_F /\alpha_U = 1.071 \pm 0.014$, and for 0.5X TBE $\alpha_F /\alpha_U = 1.113 \pm 0.048$. Our estimates are thus in reasonable agreement with earlier studies.\cite{dorfman2018odijk,alizadehheidari2015nanoconfined}

Effect due to intra-chain entanglements are ignored in our
  model. For instance,  knots\cite{micheletti2014knotting,plesa2016direct} or
  other types of  entanglements could in general give non-trivial contributions
  to the friction coefficients for linear and circular DNA.  We did, however,
  not observe any knots in our data (in   fluorescence microscopy a DNA knot is easy to discern by its higher local  intensity \cite{metzler2006diffusion}). 

Another simplification in our model is that we ignore the effect of nicking on the mechanical and frictional properties of DNA. For instance, when a nicking event occurs on only one strand, the DNA at the nick site is like a "hinge". Such a hinge is expected to change the persistence length of the DNA locally. On the other hand we may argue that since (1) the time frame of the measurements of the unfolding is shorter than the time leading up to the start of the unfolding process, and (2) the extension and the fluctuations are not changing significantly during this time, any changes in DNA mechanical properties can be neglected during the unfolding process.

Finally, note that a correction of the friction coefficient estimate should be
followed by a similar correction of the force estimate, since these are
positively correlated.

Our study demonstrates the strength of combining stochastic modeling of single molecule dynamics with the new Bayesian inference framework from Ref. \onlinecite{krog2017bayes}. This framework allow us to, not only precisely pin-point all parameters in the model with associated error estimates, but also, via its p-value, perform a check whether the model is consistent with the experimental data.

\begin{acknowledgments}
JK and MAL acknowledge the support from the Danish Council for Independent Research (grant no. 4002-00428B).
FW, BM and TA acknowledges funding from the Swedish Research Council (grant nos. 2015-5062, 2017-03865 and 2014-4305). The circular DNA was a kind gift from Alex Hastie and Denghong Zhang at BioNanoGenomics.
\end{acknowledgments}

\appendix

\section{Estimate for the unfolding force {in the extended de Gennes regime}}
\label{estimates:force}

We consider the free energy,  $G$, for a linear DNA molecule inside a nanochannel with the extension $\XU$. 
In the extended de Gennes regime\cite{Wang2011,Dai2013,Dai2014,werner2014confined}
one can apply a Flory-type mean-field argument to obtain\cite{odijk2008scaling}
\begin{equation}
\frac{G_{\rm U} }{k_BT}=A \frac{X_{\rm U}^2}{Ll_p} +B \frac{L^2 w_{\rm eff}}{X_{\rm U} D_1D_2}, 
\end{equation}
where $A$ and $B$ are number of order unity, $L$ is the contour length, $l_p$ the persistence length, and $w_{\rm eff}$ is the effective width of the molecule, while $D_1,D_2$ are the the {widths} of the nanochannel.
We treat the circular polymer as two chains of length $L/2$ which are forced to overlap, such that for an extension $X_{\rm F}$\cite{alizadehheidari2015nanoconfined},
\begin{equation}
\frac{{ G}_{\rm F}}{k_BT}=2A \frac{X_{\rm F}^2}{(L/2) l_p} +B \frac{L^2 w_{\rm eff}}{X_{\rm F} D_1D_2 }.
\end{equation}
Making the (crude) approximation $X_{\rm U} = 2X_{\rm F} = X$,  we estimate the force, see Eqs.~\ref{uforces} and~\ref{uforces2},  by
\begin{equation}
\label{f_est}
f\equiv - \frac{{ G}_{\rm F} - { G}_{\rm U}}{X/2}=\frac{2BL^2}{X^2}\frac{w_{\rm eff}}{D_1D_2}k_BT.
\end{equation}
We obtain the numbers $A$ and $B$ by exploiting the fact that analytical results are known\cite{werner2014confined} for the mean and variance of the linear extension in the extended de Gennes regime, in which case\cite{alizadehheidari2015nanoconfined}
\begin{equation}
\text{Var}(X_{\rm U}) = \frac{Ll_p}{12A} = 0.51^2Ll_p, \text{ so } A \simeq 0.32,
\end{equation}
while
\begin{equation}
\text{E}(X_{\rm U}) = L \left(\frac{Bl_pw_{\rm eff}}{2AD^2} \right)^{\frac{1}{3}} = 1.18\left(\frac{l_pw_{\rm eff}}{D^2} \right)^{\frac{1}{3}} \text{ so } B\simeq 1.05.
\end{equation}
We use the approximate values $L \simeq 43 \mu\rm m$ (the circular DNA molecules carry approximately 130 kbp), $D_1 = 100~{\rm~nm}, D_2=150~{\rm~nm}$ as well as the extension for each ionic strength $X_{0.05X} \simeq 17 \mu m$ and $X_{0.5X}\simeq 13\mu m$. 
The effective width $w_{\rm eff}$ is calculated in Ref.~\onlinecite{Iarko2015}, stating $w_{\rm eff}^{0.05X} = 26$ nm and $w_{\rm eff}^{0.5X}=10$ nm, which combine to the  crude estimates
\begin{align*}
f^{0.05X}&\simeq { 9.4 \text{ fN} }\\
f^{0.5X}&\simeq { 6.2\text{ fN}}. 
\end{align*}
From Eq.~\eqref{f_est} we see that the force should scale inversely with the $D_1$ and $D_2$ but linearly in $w_{\rm eff}$. As lower salt concentrations will yield larger effective widths\cite{persson2010dna} but only slightly smaller extension $R$, we  predict a larger force for these systems, in agreement with numerical values for the forces extracted via the Bayesian analysis, see main text.

\section{Estimate for the friction constant}
\label{estimates:gamma}

To estimate the friction coefficient, $\gamma$, we model the DNA molecule as a
solid cylinder of radius $r_1$ and the nanochannel as a surrounding cylinder
of radius $r_2$, both of infinite length. The smaller cylinder moves at a
constant speed $v_0$ along the $z$ axis, and the system is symmetric under rotations around and translations along this axis. Enforcing "no-slip" assumptions on the walls of both cylinders, we expect a nonzero gradient of velocities along the radial $r$ axis. Assuming stationarity and constant pressure along the $z$ axis, we must have 
\begin{equation}
\left(\frac{\partial^2}{\partial r^2} + \frac{1}{r}\frac{\partial}{\partial r}\right) v_z = 0
\end{equation}
with the general solution
\begin{equation}
v_z(r) = A\text{log}\frac{r}{r_2} + B.
\end{equation}
Enforcing the no-slip conditions $v_z(r_1) = v_0$ and $v_z(r_2) = 0$, we find
\begin{equation}
v_z(r) = v_0\frac{\text{log}\frac{r}{r_2}}{\text{log}\frac{r_1}{r_2}}.
\end{equation}
The viscosity $\eta$ then produces a drag per length of the cylinder $D = \gamma v_0$, which is given by the integral around the $z$ axis of the stress tensor component $\sigma_{zr} = \eta\frac{\partial v_z(r)}{\partial r}$, such that { 
$\gamma v_0 = \int_0^{2\pi r_1}d\phi\, \eta v_0/[r_1\text{log }\frac{r_1}{r_2}] = 2\pi\eta v_0/\text{log}(r_1/r_2)$. Thus, 
\begin{equation}
\label{gam_est}
\gamma   = \frac{2\pi\eta}{\text{log}\frac{r_1}{r_2}}.
\end{equation}
}
Using the viscosity of water $\eta_w \simeq 1 \text{ mPa s}$ and the ratio $r_1/r_2 = 1/50$, we find that
$$
\gamma^{\rm est} \simeq 1.6 \text{ mPa s}.
$$
From Eq.~\eqref{gam_est}, we see that there is a logarithmic dependence on the DNA cylinder width $r_1$, which should increase with lower salt concentrations, but the logarithmic form  weaken this effect.

\section{Simulations}
\label{sims}

We use Eqs.~\eqref{meansoflr} and~\eqref{yvars} to evolve the fold
conformations until either $x_l$ or $x_r$ falls below 0. The initial values
for $x_l$ or $x_r$ were estimated from the experimental kymograph. Then we use Eq.~\eqref{stage2mean} and~\eqref{stage2var} to evolve the remaining fold until this also vanishes.
Using such a simulated data set, we test the Bayesian framework by comparing the inferred values to the ones used for simulation. 

\begin{figure}[bt]
\center
\includegraphics[width = 0.35\textwidth]{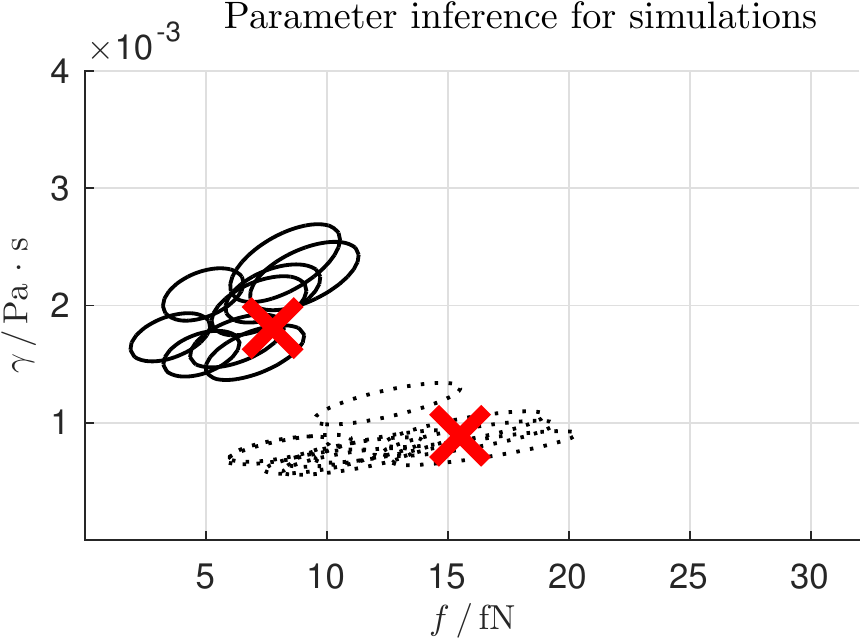}
\caption{Circles corresponding to the $1\sigma$ contours of the likelihood function $P(D|\gamma,f)$ for 20 replicated data sets as in Fig.~\ref{circles}. The parameters used for the simulations are indicated by crosses.}
\label{spreadsim}
\end{figure}

For a sample data set we use $x_l(0)$ and $x_r(0)$ along with the estimates of $\XU$, $\gamma$ and $f$ to replicate 10 artificial data sets for two different sets of parameter values, and represent the Bayesian parameter inference results in Fig.~\ref{spreadsim}.
Note how the limited statistics lead to a slight spread between the inferred values similarly to the estimates from the experimental data in Fig.~\ref{circles}.

\section{Center of mass analysis}\label{sec:com_analysis}

 Our model assumes that the center-of-mass mobility constant is
\begin{equation}
\mu = \frac{1}{\gamma \XU}.
\end{equation}
In particular, the estimate above applies to both
the motion of intact circular DNA (before unfolding) and linear DNA
(after unfolding). 
The Fokker-Planck equation for the center-of-mass coordinate takes the form
\begin{equation}
\partial_t p(x,t) = \tfrac{1}{2}\partial_x [\frac{2k_BT}{\gamma \XU}p(x,t)],
\end{equation}
which leads to the Langevin equation
\begin{equation}
\frac{dx}{dt} = \sqrt{\frac{2k_BT}{\gamma \XU}}\eta(t),
\end{equation}
with $\eta(t)\sim N(0,1)$.
For a small timestep $\Dt$, the center-of-mass coordinate thus has zero mean increments and the variance
\begin{equation}
\sigma^2 = \langle (x(t+\Dt) - x(t))^2 \rangle = \frac{2k_BT}{\gamma \XU}\Dt.
\end{equation}
For a series of $T$ measurements taken $\Dt $ apart, $\{x_i \} = x_1,x_2 \hdots x_T$, the likelihood for a specific $\gamma$ is thus
\begin{equation}
p(\{ x_i\}|\gamma,\XU) = \prod_{i=2}^T \frac{1}{\sqrt{2\sigma^2}}e^{-\frac{(x_i - x_{i-1})^2}{2\sigma^2}},
\end{equation}
where we explicify the dependence on $\gamma$ as well as $\XU$.

We found that for 0.05X TBE, there were seven trajectories which our image analysis method, see appendix \ref{image_analysis}, could provide complete edge trajectories before and after unfolding. For 0.5X, there were four such complete trajectories. We applied the Bayesian analysis to these 11 trajectories, using the likelihoods given above.

\section{Image analysis}
\label{image_analysis}

\begin{figure}[!h]
\centering
\includegraphics[width=0.4\textwidth]{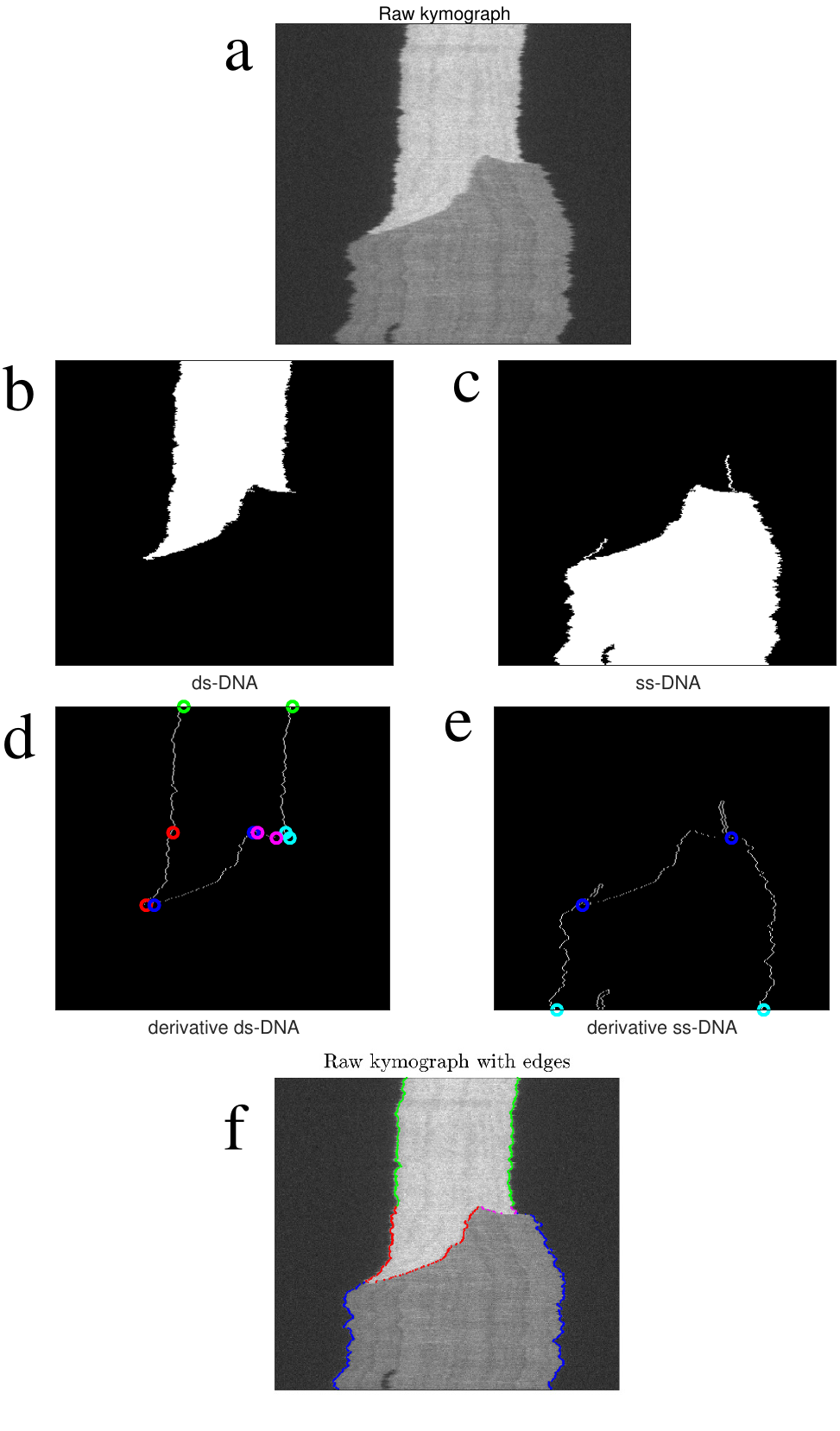}

\caption{Detecting edges in a kymograph at different times and
  linking these edges into trajectories. (a)
  Raw  kymograph. (b-c) Segmented images using the multi-Otsu method. (d-e)
  Derivative filtered version  of the images in b and c and associated keypoints. (f) Final edge
  trajectories obtained using a shortest path algorithm between keypoints in  
  the images in panels d and e. } \label{fig:edge_detection}
\end{figure}

Consider a raw kymograph, of the type in Fig. \ref{fig:edge_detection} (panel
a). The intensities in the associated image are denoted $I({\rm  row},{\rm
  col})$ for different rows and columns.  As in
Ref. \onlinecite{alizadehheidari2015nanoconfined} 
we aim to segment such a kymograph into background, linear DNA containing
pixels and circular DNA containing pixels. Also, as in
Ref. \onlinecite{alizadehheidari2015nanoconfined} we use the multi-Otsu method as a first
step (step 1 below) in this task. However, in the present study the aim is to
detect all edges at all times and connect these edges into trajectories,
rather than finding the cutting point, cutting time and unfolding time. For that reason, steps 2-4 below are new to this study. 

Our method for creating edge trajectories from a kymograph consists of four steps:
\begin{enumerate}

\item {\bf Image segmentation.}  We first segment the image into three
  regions: (i) region with two DNA double-strands (high intensity), (ii)
  region with one
  double-strand (medium intensity)  and (iii) background (low
  intensity). To that end, we use the multi-Otsu method ('multithresh' in
  Matlab) with three levels.\cite{alizadehheidari2015nanoconfined} We create two masks, one for region (i) and one for region (ii). A mask is an image which = 1 for a  pixel belonging to that region and = 0 for a region which does not belong to
  that region. We then find the largest connected component for each mask (this
  removes salt and pepper noise). Examples of segmented images for the two
  regions can be seen in Figure  \ref{fig:edge_detection} (panels b and c).

\item {\bf Initial edge detection.}  We next run a median filter with a 3$\times$3
  window on 
  the each of the two masks from step 2 in order to correct for corrupt edges. We then apply
  a derivative filter along rows,  i.e., identify all 0-1 (left edge) and 1-0
  (right edge)  transitions for all rows. Ideally, we now have all edges
  identified -- for the unfolding regime, we should have four edges
  and for the intact  circular DNA and  fully unfolded DNA we should have two edges for
  all time frames. In practice, it could happen that the derivative filter
  provides more than the  expected number of edges. This sensitivity-to-noise issue is dealt with next.

\item {\bf Keypoint detection.} To deal with the problem above, we detect
  robust ``keypoints'' in the masks for region (i) and (ii), see
  Fig. \ref{fig:edge_detection} (panels d and e). There are  several keypoints: two start edges -- the leftmost  edge and rightmost edges  in the first time frame for region (i). Two stop edges -- the  leftmost and rightmost edges in the last time point in region (ii). The cutting point is the point at which the cut occurred and is identified as the first time frame where four edges were detected in the mask image for region (i). The leftmost edge
  and rightmost edge at the cutting time are also keypoints. Finally, we identify the time
  at which the first and second fold disappeared in the
  mask for region (i). The left and right fold positions at these times
  serve as keypoints. The keypoints as defined above are robust
  and not sensitive  to noise for the data at hand. 
 
\item {\bf Edge tracing}. With the keypoints above as input, we then run a
  shortest path algorithm  between these points in order to determine the edge
  trajectories. To that end, we consider each of the white pixels in the
  derivative filtered images (panels d and e in Figure
\ref{fig:edge_detection}) as a node in a graph. We connect each node to a node in the next row if it is  within {\it colSpan} number of pixels. We
  then find the path from the start keypoint to the end keypoint which
  minimizes a cost function using Dijkstra's shortest path algorithm
  ('shortestpath' in Matlab). The cost function is a sum over all rows where a  jump in the horizontal direction is given a weight $w^n$, where $n$ is
  difference in row numbers between time frames. We choose, {\it colSpan}=17
  and $w$=3. 

\end{enumerate}

An example of the final edge trajectories is found in Fig.
\ref{fig:edge_detection} (panel f).

\end{document}